\documentclass[%
 aip,
 amsmath,amssymb,
 reprint,%
]{revtex4-1}

\usepackage{graphicx}
\usepackage{dcolumn}
\usepackage{bm}

\usepackage[utf8]{inputenc}
\usepackage[T1]{fontenc}
\usepackage{mathptmx}
\usepackage{etoolbox}
\usepackage{amsmath}
\usepackage{braket}

\makeatletter

\def\@email#1#2{%
 \endgroup
 \patchcmd{\titleblock@produce}
  {\frontmatter@RRAPformat}
  {\frontmatter@RRAPformat{\produce@RRAP{*#1\href{mailto:#2}{#2}}}\frontmatter@RRAPformat}
  {}{}
}%
\makeatother
\begin{document}

\title{Cooperative Conformational Transitions in Macromolecules under Mechanical Stretching. An Exactly Solved Model for Single Molecule Experiments} 

\author{Javier Orradre}
\email{fmas@ub.edu}
\affiliation{Department of Materials Science and Physical Chemistry \& Institute of Theoretical and Computational Chemistry (IQTC), University of Barcelona, Barcelona, Catalonia, Spain.}

\author{Pablo M. Blanco}%
\affiliation{Department of Chemistry, Physics and Environmental and Soil Sciences \& Agrotecnio, University of Lleida, Lleida, Catalonia, Spain.}

\author{Sergio Madurga}
\affiliation{Department of Materials Science and Physical Chemistry \& Institute of Theoretical and Computational Chemistry (IQTC), University of Barcelona, Barcelona, Catalonia, Spain.}

\author{Marina I. Giannotti}
\affiliation{Department of Materials Science and Physical Chemistry \& Institute of Theoretical and Computational Chemistry (IQTC), University of Barcelona, Barcelona, Catalonia, Spain.}
\affiliation{Nanoprobes \& Nanoswitches group, Institute for Bioengineering of Catalunya (IBEC), The Barcelona Institute of Science and Technology (BIST), Barcelona, Catalonia, Spain.}
\affiliation{CIBER-BBN, ISCIII, Barcelona, Catalonia, Spain.}

\author{Francesc Mas}
\affiliation{Department of Materials Science and Physical Chemistry \& Institute of Theoretical and Computational Chemistry (IQTC), University of Barcelona, Barcelona, Catalonia, Spain.}

\author{Josep Lluís Garcés}
\affiliation{Department of Chemistry, Physics and Environmental and Soil Sciences \& Agrotecnio, University of Lleida, Lleida, Catalonia, Spain.}

\date{\today}

\begin{abstract}
The stretching behavior of linear macromolecules undergoing conformational transitions is investigated. 
An exact solution is provided for a two-state system within the elastic freely jointed chain  model. 
This minimal framework contains the smallest set of parameters required to describe such transitions: two Kuhn lengths, two elastic force constants, a free energy difference between both states and a nearest-neighbor interaction energy accounting for cooperativity. 
Explicit analytical expressions are derived for the chain extension and the probabilities of each state as functions of the applied force. 
The approach accurately reproduces the experimental force–extension curves of poly(ethylene-glycol) (PEG) and hyaluronic acid (HA), revealing no cooperativity for PEG and negative cooperativity for HA. 
It also describes the B-DNA to S-DNA conformational transition, a process that exhibits positive cooperativity. 
We analyze the mathematical conditions required for a transition and identify two fundamental driving mechanisms: differences in Kuhn lengths and differences in force constants. 
Extensions of the model to systems with more than two conformational states per Kuhn segment are also discussed. 
The results presented here apply equally to transitions that are intrinsic to the macromolecular structure or induced by ligand–receptor interactions, unifying both cases within a single thermodynamically consistent framework.   

\end{abstract}

\maketitle
\begin{figure*}
\includegraphics[scale=0.9]{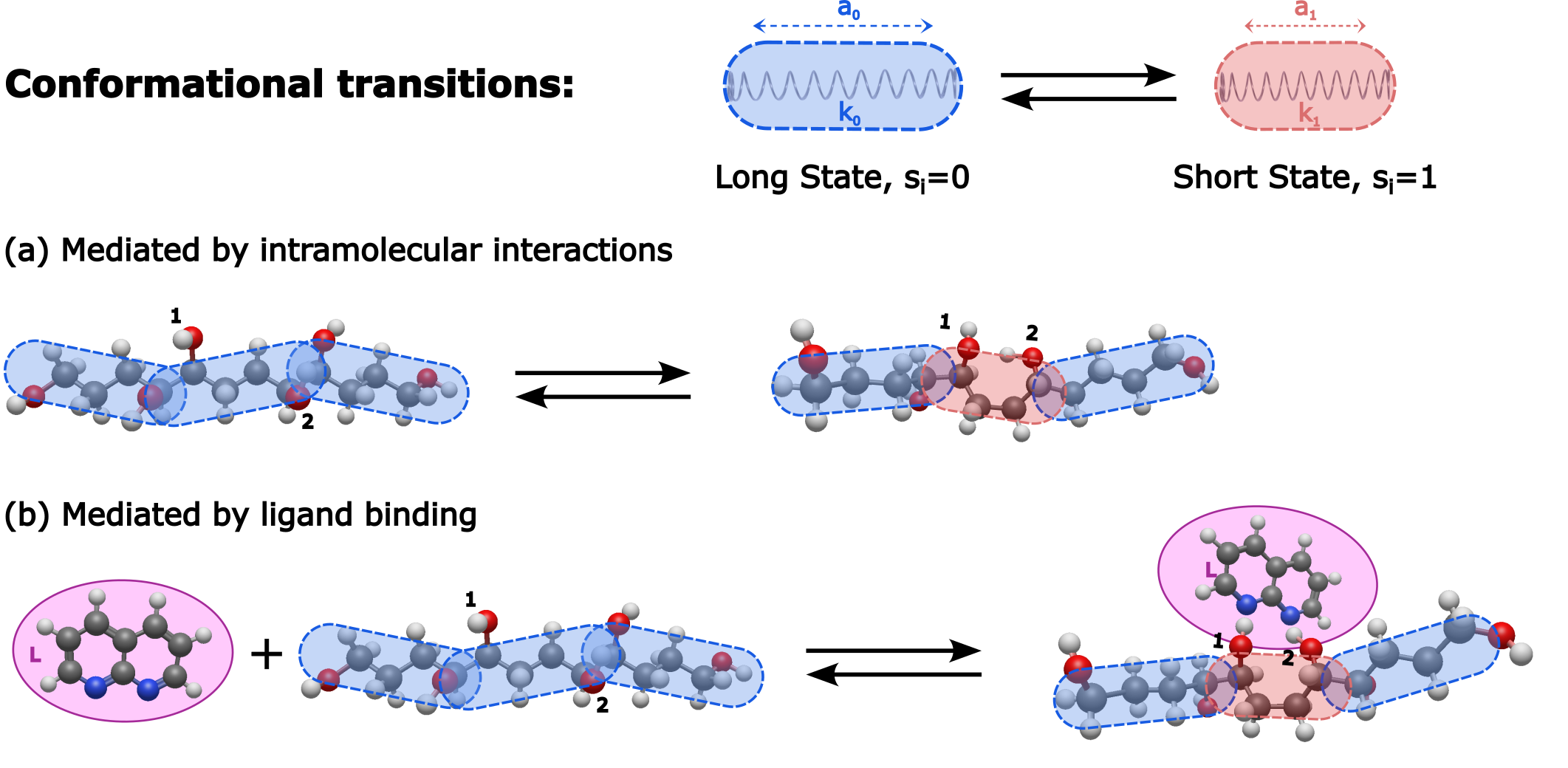}
\caption{In the two-state model, the chain is described as a sequence of fragments, each representing a Kuhn length, i.e., the minimum distance over which orientational correlations are lost. Each fragment can adopt one of two possible conformational states, characterized by their rest lengths $a_0$ (long state) and $a_1$ (short state), and their corresponding elastic constants, $k_0$ and $k_1$. Transitions between these two states can occur via two distinct mechanisms: (a) intramolecular interactions, such as hydrogen bonding, hydrophobic interactions, or steric hindrance; or (b) ligand binding to the chain fragments. Both mechanisms can be described within equivalent thermodynamic frameworks.}
\label{fig:modelos}
\end{figure*}

\section{\label{sec:Introduction}Introduction}

The study of mechanically induced chemical and conformational transitions has experienced significant progress in the past decades, largely driven by the development of single-molecule force spectroscopy techniques such as atomic force microscopy (AFM) and magnetic/optical tweezers.\cite{Giannotti2007a,Camunas-Soler2016,Strick2003} 
Once the force-extension curves are obtained, they can, in principle, be analyzed in terms of the underlying physicochemical events. 
For example, Wiita \textit{et al.} found that the catalytic activity of thioredoxin can be regulated by applying an external force to its substrate.\cite{Wiita2007} Surface desorption of polypeptides,\cite{Friedsam2006} \textit{cis}-to-\textit{trans} isomerization of carbon-carbon double bonds, boat-to-chair conformational transitions,\cite{Marszalek1998,Radiom2016} pinpointing of mechanochemical bond rupture using embedded macrocycles \cite{Schutze2015} and hydration and solvent effects \cite{Liese2017,Radiom2017b} can all be triggered by macromolecular stretching. 
In all these cases, chemical events produce mechanical work, or mechanical forces can drive chemical transformations, thereby paving the way for new nanotechnologies based on mechanochemical properties.

The stretching properties of polymer chains are strongly coupled with conformational transitions,\cite{Marko1997,Garces2006,Radiom2017a,Blanco2019a,Blanco2023} which can occur through two main mechanisms, outlined in Fig. \ref{fig:modelos}. 
On the one hand, they may arise from changes in the relative population of two or more conformational states. Examples include the stretching-induced transition from B-DNA to S-DNA \cite{Ahsan1998} and the breaking of hydrogen bonds in poly(ethylene-glycol).\cite{Oesterhelt1999,Liese2017} 
On the other hand, conformational transitions can also result from the binding of an external ligand.\cite{Wiita2007,Zhang2008,Krysiak2014,Jarillo2017} 
For instance, IHF (integration host factor), RecA protein and chromatin are three examples of ligands that bind to double-stranded DNA, which strongly modify its stretching properties.
\cite{Marko1997,Leger1998,Ali2001} As will be shown below, both mechanisms can be treated within formally equivalent thermodynamic frameworks.

Nevertheless, models that explicitly incorporate stretching-induced conformational transitions are rarely found in the literature. 
The two most widely used models for polymer stretching, largely due to their simplicity, are the Freely Jointed Chain (FJC) and the Worm-Like Chain (WLC).\cite{Flory1969,Marko1995,Bustamante1994,Giannotti2007a} 
In the FJC model, the polymer is represented as a series of fragments with uncorrelated relative orientations. 
The FJC model can be properly extended to include chain elasticity, resulting in the Elastic Freely Jointed Chain (EFJC) model.\cite{Balabaev2009,Radiom2017a}
In contrast, in the WLC model the chain is treated as a semiflexible filament whose bending rigidity determines its persistence length. 
In both cases, the internal conformational degrees of freedom are neglected, and conformational transitions are not accounted for. 
To overcome the limitations of the WLC and FJC models, Oesterhelt \textit{et al.}  proposed a heuristic two-state conformational model which successfully fitted the force-extension curves of PEG.\cite{Oesterhelt1999} 
However, as pointed out previously by Radiom and Borkovec,\cite{Radiom2017a} this model is not thermodynamically consistent since Maxwell relations are not satisfied.\cite{Zhang2008} 
Marko and Siggia also proposed a similar model to characterize the binding of proteins to DNA.\cite{Marko1997}
In all aforementioned models, the conformational states are assumed to be uncorrelated along the chain. 

However, conformational transitions are generally cooperative phenomena,\cite{Garces2006} and therefore suitable models should incorporate an interaction energy between at least neighboring conformational states. 
In a relatively recent work, Radiom and Borkovec proposed a ligand-receptor extension of the EFJC model which accounts for the nearest neighbor interactions treated within the mean-field approximation.\cite{Radiom2017a} 
The chain is divided into fragments which can adopt two possible states, empty or occupied, with different energy and length. 
Two consecutive occupied sites interact through an additional energy term that accounts for cooperativity. 
This model can be regarded as minimal, in the sense that it contains the smallest set of parameters required to describe two-state cooperative conformational transitions within the EFJC approximation: the Kuhn lengths for the occupied and empty states, a free energy difference between them, and the interaction energy accounting for cooperativity. Shortly afterwards, Benedito and Giordano\cite{Benedito2018} proposed an exact solution to this model for finite chains using transfer matrix methods, a common and useful tool in statistical mechanics \cite{Flory1969,Chandler1987}. A similar framework, but within the WLC approximation, was introduced by Cluzel \textit{et al.}  \cite{Cluzel1996} and later refined by Ahsan \textit{et al.}  \cite{Ahsan1998} to explain the B-DNA to S-DNA conformational transition.

In this work, we provide an alternative and simpler derivation\cite{Benedito2018} specific for the $N\rightarrow\infty$ case, suitable for describing stretching experiments. We also verify that the exact treatment eliminates the spurious phase transitions inherent to mean field approaches and how it yields closed analytical expressions for the macromolecule key properties. The  corresponding derivations of the model are outlined in Section \ref{sec: EXACT SOLUTION FOR THE TWO-STATE FREELY JOINTED CHAIN}. 
In Section \ref{sec:experiments}, we show that the model successfully explains the reported experimental force–extension curves of PEG, HA and DNA. 
The necessary mathematical conditions for a transition to occur are analyzed in section \ref{sec: transitions}. 
Finally, in Section \ref{sec: Generalization to more than two conformational states}, we discuss the generalization of the method to systems with more than two conformational states or to cases where two ligands compete for the same receptor.

\section{\label{sec: EXACT SOLUTION FOR THE TWO-STATE FREELY JOINTED CHAIN}ALTERNATIVE DERIVATION FOR THE EXACT SOLUTION OF THE TWO-STATE ELASTIC FREELY JOINTED CHAIN}

Consider an Elastic Freely Jointed Chain (EFJC) composed of $N$ fragments whose orientation and length are characterized by a set of vectors $\{\mathbf{b}_{i}\}$. 
Each fragment represents a Kuhn length, which corresponds to the minimum distance required for orientational correlations to be lost. 
Each fragment can exist in two possible conformational states, denoted by the binary state variables $s_i$, taking values 0 or 1. 
If the conformational change is induced by ligand binding (case b in Fig. \ref{fig:modelos}), $s_i$ also represents the binding state of the fragment, i.e., empty ($s_i=0$) or occupied ($s_i=1$).

\begin{figure*}[t]
    \centering
    \includegraphics[width=1\linewidth]{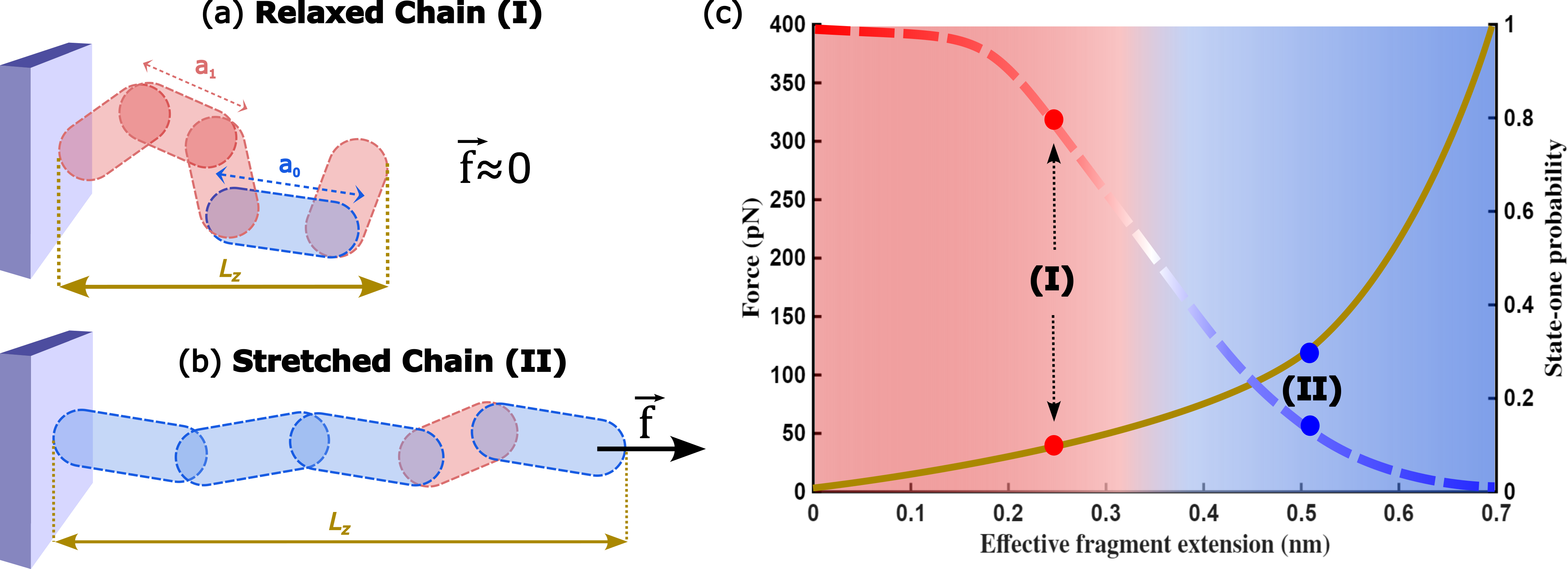}
    \caption{Schematic representation of two possible configurations corresponding to a relaxed (a) and a stretched (b) chain composed of $N$ elastic fragments. At low forces, the fragment orientations are uncorrelated, corresponding to a random coil behavior. Each fragment can exist in two states, denoted as 0 (blue) and 1 (red), with rest lengths $a_0$ and $a_1$ and elastic constants $k_0$ and $k_1$, respectively. $L_z$ denotes the projection of the end-to-end vector along the direction of the applied force $f$, and we define $x=L_z/N$ as the effective fragment extension. Panel (c) shows the applied force, $f$ (solid line), and the probability of state 1, $\theta$ (dashed line), as functions of the effective fragment extension, $x$.} 
    \label{fig:theoretical extension}
\end{figure*}

In this way, the chain conformation is specified by the set $\{s_i, \mathbf{b}_i\}$ with $i=1,\dots N$. 
In order to account for cooperative effects, an additional interaction energy $\varepsilon$ is incorporated when two neighboring fragments are in state $s_i = 1$. 
The free energy of a particular chain configuration reads
\begin{gather}
    \nonumber F\left(\{s_i, \mathbf{b}_i\}\right)=-\sum_{i=1}^N \mathbf{f}\cdot\mathbf{b}_i\;+\\\nonumber +\sum_{i=1}^N \left[\frac{k_0}{2}\left(b_{i}-a_0\right)^2(1-s_i)+\frac{k_1}{2}\left(b_{i}-a_1\right)^2s_i\right] + \\
    + \sum_{i=1}^N \mu\,s_i + \sum_{i=1}^{N-1} \varepsilon\,s_i s_{i+1}
    \label{eq: global elastic free energy}
\end{gather}
where $\mathbf{f}$ is the force vector, which is assumed to act in the $z$-direction. 
The term $\mu$ represents the free energy difference between the two different conformational states. 
In the case that the conformational changes are induced by ligand binding, $\mu$ can also be interpreted as the reduced chemical potential  so that $\mu = -\beta^{-1}\ln\left(K a_{\text{Lig}}\right)$ where $K$ is the intrinsic binding constant of the receptor, $\beta = 1/(k_{\mathrm{B}}T)$ is the inverse of the thermal energy and $a_{\text{Lig}}$ is the ligand activity.\cite{Blanco2018b,Blanco2019b}
Note that the treatment of the two physical situations is formally equivalent, only the interpretation of $\mu$ changes. 
Finally, the term between brackets accounts for the elastic energy of the fragments, which are treated as harmonic springs. 
Each conformational state has its own rest length ($a_{0}$ and $a_{1}$) and elastic constant ($k_{0}$ and $k_{1}$), as depicted in Fig.\,\ref{fig:modelos}.

The partition function at fixed temperature is defined by 
\begin{gather}
    \Xi \left(f,\mu\right)=\sum_{\{s_i\}}\int\dots\int \{d\mathbf{b}_i\} e^{-\beta F\left(\{s_i, \mathbf{b}_i\}\right)}
    \label{eq: global part func}
\end{gather}
where the sum and integral are taken over the possible conformational states $\{s_i\}$ and fragment vectors $\{\mathbf{b}_i\}$, respectively. Observe that integration over the scalar product removes the vectorial character of the force, so that the partition function is determined solely by its magnitude, $f$.
The two key quantities are the effective extension per fragment, $x=L_z/N$ (where $L_z$ is the total chain extension along the force direction), and the fraction of fragments in state $s_{i}=1$, $\theta$ (state-one probability), outlined in Fig.\,\ref{fig:theoretical extension}. 
In the ligand-receptor literature, $\theta$ is commonly referred to as the occupation degree. 
Both quantities can be derived from the partition function as
\begin{align}
\theta(f,\mu) &= -\frac{1}{N\beta} \left(\frac{\partial \ln \Xi}{\partial \mu}\right)_f, &
x(f,\mu) &= \frac{1}{N\beta} \left(\frac{\partial \ln \Xi}{\partial f}\right)_\mu
\label{eq:theta and Lz generals}
\end{align}
and automatically satisfy thermodynamic consistency, given by the Maxwell relation
\begin{gather}
    \left(\frac{\partial\theta}{\partial f}\right)_\mu=-\left(\frac{\partial x}{\partial \mu}\right)_f.
    \label{eq:Mawxell Thermod consistency}
\end{gather}

Since each summand in Eq.\,\ref{eq: global elastic free energy} contains only one vector $\mathbf{b}_i$, the integration in Eq.\,\ref{eq: global part func} can be factorized. Using now the identity\cite{Garces2006,Orradre2024} 
\begin{gather}
    \int d\mathbf{b_i}\,e^{-\beta \left[F_0\left(\mathbf{b_i}\right)\left(1-s_i\right)+F_1\left(\mathbf{b_i}\right)s_i\right]} =e^{-\beta \left[g_0\left(1-s_i\right)+g_1 s_i\right]}
    \label{eq: conformational g's def}
\end{gather}
where $F_{\alpha}\left(\mathbf{b}_i\right)$ is the single-fragment conformational free energy with vector $\mathbf{b}_i$ in state $\alpha=0,1$

\begin{gather}
F_{\alpha}\left(\mathbf{b_i}\right)=\mathbf{f}\cdot\mathbf{b_i}+\frac{k_\alpha}{2}\left(b_i-a_\alpha\right)^2
    \label{eq: signle-fragment free energy}
\end{gather}
and  
\begin{gather}
    g_\alpha(f)=-\beta^{-1}\ln q_\alpha(f),
    \label{eq:g_alpha definition}
\end{gather}
where $q_\alpha$ is the single-fragment partition function defined as
\begin{gather}
    q_\alpha(f) = \int d\mathbf{b} e^{-F_{\alpha}\left(\mathbf{b}\right)}
    \label{eq: single-fragment partition function}
\end{gather}
which, in spherical coordinates, reads
\begin{gather}
    q_\alpha=\int_0^{2\pi}d\phi\int_0^\infty\int_{-1}^1 db\;d\cos\vartheta e^{-\beta f b \cos\vartheta-\frac{\beta k_\alpha}{2}\left(b-a_\alpha\right)^2}=\nonumber\\=
    4\pi\int_0^\infty db \frac{\sinh \left(\beta b f\right)}{\beta b f} e^{-\frac{\beta k_\alpha}{2}\left(b-a_\alpha\right)^2}.
    \label{eq: qalpha FEJC elastic bonds}
\end{gather}

For fragments stiff enough such that $\sqrt{\beta k_\alpha a_\alpha^2}\gg1$, the integral in Eq.\,\ref{eq: qalpha FEJC elastic bonds} can be evaluated explicitly \cite{Balabaev2009,Radiom2017a}

\begin{gather}
    q_\alpha=\left(1+\frac{1}{\beta k_\alpha a_\alpha^2}\right)^{-1}\left[\frac{\sinh\left(\beta a_\alpha f\right)}{\beta a_\alpha f}+\frac{\cosh\left(\beta a_\alpha f\right)}{\beta k_\alpha a_\alpha^2}\right]e^{\beta f^2/(2k_\alpha)}
    \label{eq: qalpha FEJC elastic bonds approximation}
\end{gather}
In Eq.\,\ref{eq: qalpha FEJC elastic bonds approximation}, $q_\alpha$ has been renormalized so that $q_\alpha\rightarrow1$ and $g_\alpha\rightarrow0$ when $f\rightarrow0$. This redefinition has no physical consequences but ensures that at zero force the system's free energy is solely determined by $\mu$ and $\varepsilon$, which will be convenient in the following.

Inserting Eq.\,\ref{eq: conformational g's def} into Eq.\,\ref{eq: global part func}, the partition function can now be expressed in terms of an effective free energy that depends only on the variables $s_i$
\begin{gather}
    \Xi =\sum_{\{s_i\}}e^{-\beta F_{\mathrm{eff}}\left(\{s_i\}\right)}
    \label{eq: global effective part func}
\end{gather}
with
\begin{gather}
    F_{\mathrm{eff}}\left(\{s_i\}\right)=
    Ng_0+\widetilde{\mu}\;\sum_i^N s_i+\varepsilon \sum_i^{N-1} s_i s_{i+1}
    \label{eq:global effective free energy}
\end{gather}
where $\widetilde{\mu}$ is the force-dependent transition free energy
\begin{gather}
\widetilde{\mu}\left(f,\mu\right)=g_1\left(f\right)-g_0\left(f\right)+\mu
    \label{eq:effective chemical potential}
\end{gather}
\begin{figure}[b]
    \centering
    \includegraphics[width=1\linewidth]{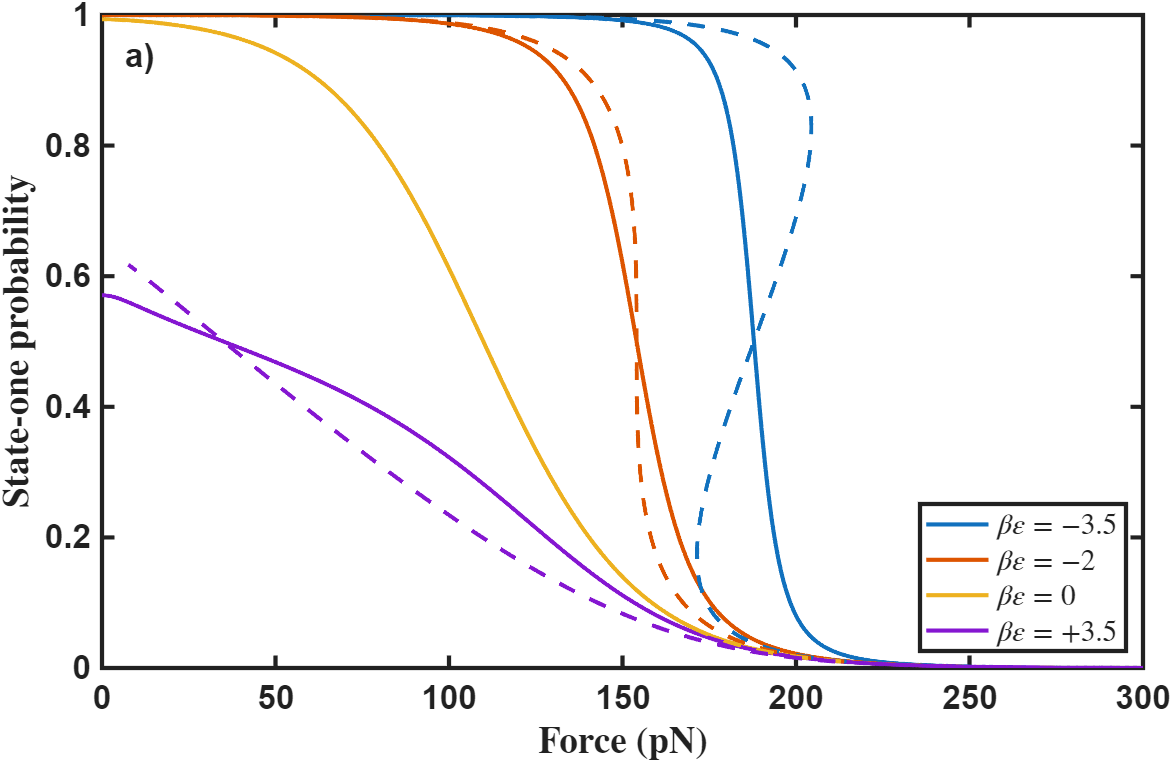}
    \includegraphics[width=1\linewidth]{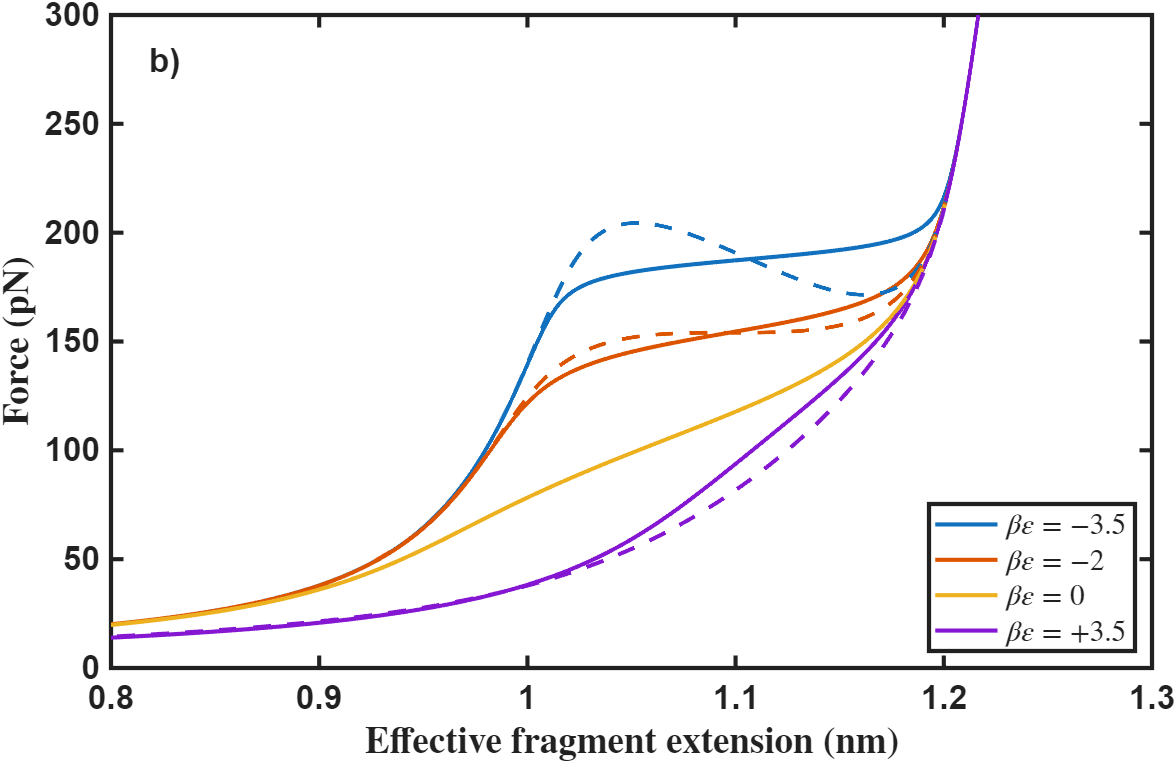}
    \caption{Comparison between the exact (solid lines) and the mean field (dashed lines) solutions for (a) the state-one probability $\theta$ \textit{vs} force $f$ and (b) the force $f$ \textit{vs} the effective fragment extension $x$ (b) of the two-states EFJC model. The parameters are $T=298$\,K\, $\beta\mu=-5$, $a_0=1.2$\,nm, $a_1=1.0$\,nm, $k_0=10\,\mathrm{N/m}$ and and $k_1=5\,\mathrm{N/m}$. Interaction energies vary in the range $\beta\varepsilon=(-3.5 , +3.5)$.}
    \label{fig:comparison MF}
\end{figure}
The free energy from Eq.\,\ref{eq:global effective free energy} is equivalent to that of a one-dimensional Ising model with nearest-neighbor interactions where the energy values depend on the exerted force.\cite{Koper2001,Garces2006} 
An elegant way to calculate the partition function is by using transfer matrices, a technique commonly used in statistical mechanics.\cite{Flory1969,Chandler1987}
As will be shown later, this approach can be easily generalized to more than two conformational states (or, equivalently, to more than two ligands in the ligand–receptor framework). 
Detailed calculations are available elsewhere \cite{Koper2001,Garces2006,Benedito2018} or in the Supporting Information (SI) section S-1. 
Here we provide the final result 
\begin{equation}
\Xi=q_0^{N}\left(\mathrm{\mathbf{v_i}}\mathbf{\widetilde{T}}^{N}\mathrm{\mathbf{v_f^T}}\right)\quad\mathrm{with}\quad\mathbf{\mathbf{\widetilde{T}}}=\left(\begin{array}{cc}
1 & \tilde{z}\\
1 & \tilde{z}u
\end{array}\right)
\label{eq:pt single ligand effective}
\end{equation}
where
\begin{equation}
\tilde{z}\left(f,\mu\right)=\frac{q_1\left(f\right)}{q_0\left(f\right)}z
\label{eq:effective z general}
\end{equation}
where $z=\exp\left(-\beta\mu\right)$ and $u=\exp\left(-\beta\varepsilon\right)$ are the Boltzmann factors corresponding to $\mu$ and $\varepsilon$. 
The initial $\mathrm{\mathbf{v_i}}=\left(1,0\right)$ and final $\mathrm{\mathbf{v_f}}=\left(1,1\right)$ vectors are required to account for the non-periodic boundary conditions.\cite{Koper2001} 
In the ligand-receptor picture, the term $q_0^N$ corresponds to the partition function of the empty chain, while $\mathbf{\widetilde{T}}$ incorporates ligand binding with the force-dependent binding constant $\widetilde{K}\left(f\right)=q_1(f)/q_0(f)\;K$. 
For long chains, we can take the limit $N\rightarrow\infty$ and Eq.\,\ref{eq:pt single ligand effective} then leads to

\begin{equation}
    \Xi\sim\left(q_0\lambda_\mathrm{max}\right)^N
    \label{eq: part func infinity}
\end{equation}
where $\lambda_\mathrm{max}$ is the largest eigenvalue of $\tilde{\mathbf{T}}$, given in this case by\cite{Koper2001}
\begin{equation}
\lambda_{\mathrm{max}}\left(f,\mu\right)=\left(\frac{1+\tilde{z}u}{2}\right)+\sqrt{\tilde{z}+\left(\frac{1-\tilde{z}u}{2}\right)^{2}}\label{eq:VAP max}
\end{equation}
Inserting Eq.\,\ref{eq: part func infinity} into Eq.\,\ref{eq:theta and Lz generals} yields the state-one probability
\begin{equation}
\theta\left(f,\mu\right)=\left[2+\left(\frac{\lambda_{\mathrm{max}}}{\tilde{z}}\right)\left(\frac{1-\tilde{z}u}{1-u+\lambda_{\mathrm{max}} u}\right)\right]^{-1}
\label{eq: theta FJEC}
\end{equation}
and the effective fragment extension
\begin{equation}
x\left(f,\mu\right)=x_0\left(1-\theta\right)+x_1\theta
\label{eq: x para FJEC single ligand}
\end{equation}
where
\begin{equation}
x_\alpha\left(f\right)=a_\alpha\left[\frac{\coth\left(\beta fa_\alpha\right)-\frac{1}{\beta fa_\alpha}+\frac{f}{k_\alpha a_\alpha}}{1+\frac{f}{k_\alpha a_\alpha}\coth\left(\beta fa_\alpha\right)}+\frac{f}{k_\alpha a_\alpha}\right]
\label{eq: length per fragment FEJC}
\end{equation}
represents the fragment extension in the force direction in the state $\alpha=0,1$. To summarize, Eq.\,\ref{eq: theta FJEC} together with Eqs.\,\ref{eq: x para FJEC single ligand} and \,\ref{eq: length per fragment FEJC}, form the set of equations for the exact solution\cite{Benedito2018} of the two-state EFJC model in the large chain limit ($N\rightarrow\infty$).

The mean-field version of Eq.\,\ref{eq: theta FJEC} corresponds to the well-known Frumkin isotherm
\begin{gather}
\theta=\frac{\tilde{z}\exp{\left(-2\beta\varepsilon\theta\right)}}{1+\tilde{z}\exp{\left(-2\beta\varepsilon\theta\right)}}
\label{eq: theta Radiom}
\end{gather}
which was used to solve the approximation proposed in Radiom and Borkovec.\,\cite{Radiom2017a} Under this mean-field approach, each fragment interacts with the average state-one fragment probability of its neighbors, so local correlations are neglected. 
Note that Eq.\,\ref{eq: theta Radiom}, unlike Eq.\,\ref{eq: theta FJEC}, is an implicit equation for $\theta$. 

Fig.\,\ref{fig:comparison MF} shows the state-one probability and the effective fragment extension for the exact solution (Eq.\,\ref{eq: theta FJEC}) and the mean-field approximation (Eq.\,\ref{eq: theta Radiom}). 
Within the mean-field approximation, both quantities become multi-valued functions of the force, which can be interpreted as spurious thermodynamic phase transitions. 
However, it is a well-established result in statistical mechanics that true phase transitions are forbidden in one-dimensional systems with short-range interactions.\cite{Chandler1987}
These artifacts are absent in the exact solution. Nevertheless, the conformational transitions can still be very sharp for certain parameter combinations, as illustrated in the figure. 
The conditions under which this occurs are analyzed in detail in Section \ref{sec: transitions}.

\section{\label{sec:experiments}INTERPRETATION OF EXPERIMENTAL FORCE-EXTENSION CURVES}

In this section, we use the two-state EFJC model to analyze reported experimental force–extension curves of  poly(ethylene-glycol) (PEG)\cite{Oesterhelt1999} and hyaluronic acid (HA)\cite{Giannotti2007b} obtained by AFM-based single molecule force spectroscopy (AFM-SMFS). 
For completeness, and given the paradigmatic role of DNA in stretching studies, we also apply the model to the B-DNA to S-DNA transition reported by Cluzel \textit{et al.} \cite{Cluzel1996} Typically, in AFM-based reported experiments,\cite{Marszalek1998,Li1999,Oesterhelt1999,Marszalek2003,Giannotti2007b,Yu2024} a large number of force–extension measurements are performed on polymer chains which, due to sample polydispersity and the \textit{random attachment} problem,\cite{Farrance2015,Liu2006,Liu2007} correspond to molecules with unknown contour lengths. To make the experiments comparable, the extension values corresponding to the different contour lengths are rescaled by the extension measured at an arbitrarily selected reference force, $L_z(f_\mathrm{norm})$. In this way, the rescaled force–extension curves, collapse onto a single master curve of $f$ versus $L_z^{\mathrm{norm}}$.

This normalized extension, $L_z^{\mathrm{norm}}$, is proportional to the effective fragment extension
\begin{equation}
    L_z^{\mathrm{norm}}(f) = L_z(f)/L_z(f_\mathrm{norm})=Cx(f)
\label{eq: normalized L experimental}
\end{equation}where $C=x(f_\mathrm{norm})^{-1}$ is the inverse of the effective relative extension at the normalization force $f_\mathrm{norm}$ (the complete derivation for the constant $C$ is provided as SI in section S-3). 
The parameters of the two-state model, are obtained by fitting Eq.\,\ref{eq: normalized L experimental} in combination with Eqs.\,\ref{eq: theta FJEC} and \ref{eq: x para FJEC single ligand} to the experimental master curve. The fitting and error estimation have been performed using the Levenberg–Marquardt non-linear regression algorithm.\cite{Levenberg1944,Marquardt1963}

\begin{figure}[b]
  \centering
  \includegraphics[width=\linewidth]{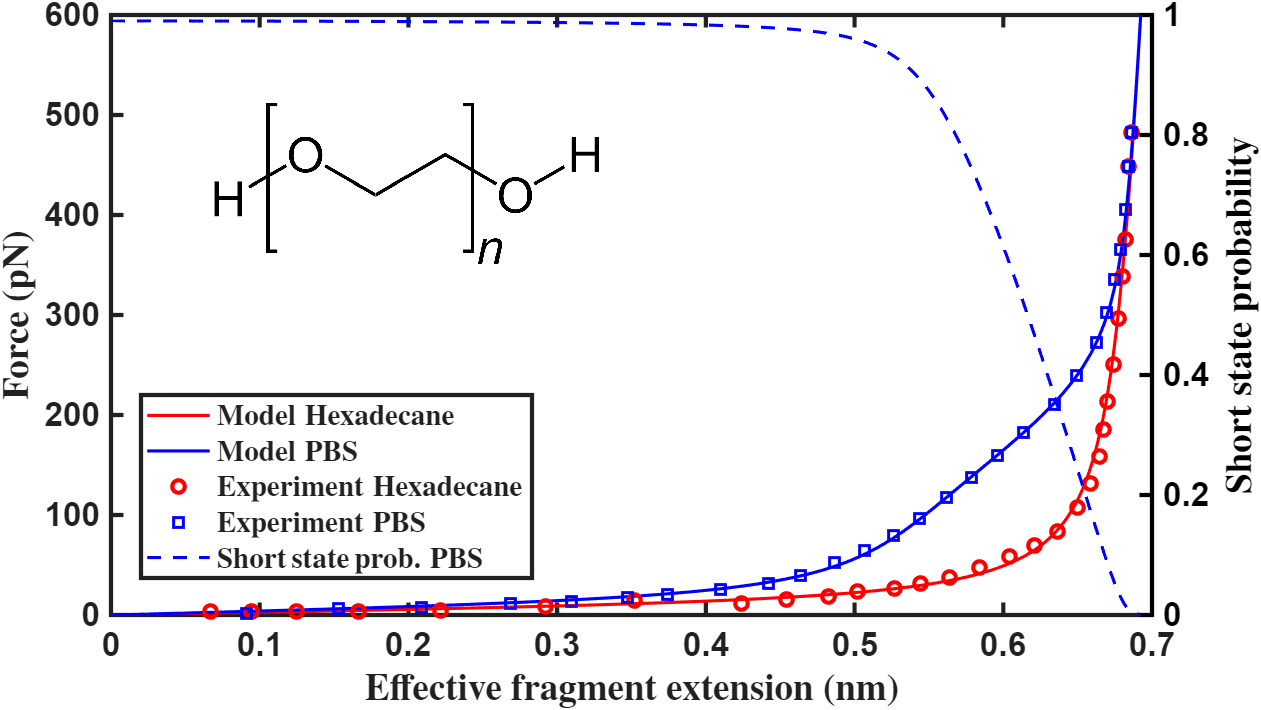}
  \caption{Experimental force-extension curves of PEG in phosphate-buffered saline (PBS) at pH 7.4 and 130 mM NaCl (blue squares) and hexadecane (red circles) at $T=25$\textdegree C from Ref.\cite{Oesterhelt1999} together with the best fits from the two-state EFJC model (solid lines). 
  In hexadecane, the conventional single-state EFJC model provides an excellent fit, yielding a Kuhn length of $a_\mathrm{L}=(6.8\pm0.3)\,$\AA$\;$ and a force constant $k_\mathrm{L} = (4 \pm 2)\times 10^{1}\,\mathrm{N/m}$. In contrast, in PBS an additional short state with Kuhn length $a_\mathrm{S}=(5.58\pm0.04)$ and elastic constant $k_\mathrm{S}=(6\pm3)\,\mathrm{N/m}$ is required to reproduce the data. The short state arises from the formation of intramolecular hydrogen bonds which are hindered in hexadecane. Consequently a conformational transition takes place when the short conformation is disrupted due to molecular stretching. The best-fit conformational free energy of the long state is $\beta\mu=-(5.0\pm0.4)$ indicating that the short conformation is significantly more stable at low forces. The obtained interaction energy is very small $\beta\varepsilon_\mathrm{SS}=(0.2\pm0.4)$, revealing that no cooperativity is present. The blue dashed line corresponds to the short state probability, $\theta$.}
  \label{fig: PEG LevMarq25}
\end{figure}

For an easier interpretation of the experiments in terms of the model, in this section the state notation ‘0’ and ‘1’ will be replaced by L (‘long’) and S (‘short’) respectively, and the interaction energy $\varepsilon$ will be redefined as $\varepsilon_\mathrm{SS}$, as the short fragments are the interacting ones. 

\subsection{Non-cooperative stretching ($\varepsilon_\mathrm{SS} \approx 0$): poly(ethylene-glycol) (PEG)}

Because of its simple structure and chemical composition, the stretching behavior of poly(ethylene-glycol) has been extensively studied both theoretically and experimentally.\cite{Oesterhelt1999,Gaballa2023,Kolberg2019,Livadaru2003b,Heymann1999,Kreuzer2001c,Begum1997} In Fig.\,\ref{fig: PEG LevMarq25}, we plot the experimental force–extension curves from Oesterhelt \textit{et al.},\cite{Oesterhelt1999} obtained in hexadecane (red circles) and in PBS (blue squares), together with the best fit of the two-state model.

In hexadecane, the conventional single-state EFJC model (red solid line) provides an excellent fit to the experimental data, yielding a Kuhn length of $a_\mathrm{L} = (6.8 \pm 0.3)\,\text{\AA}$ and a force constant $k_\mathrm{L} = (4 \pm 2)\times 10^{1}\,\mathrm{N/m}$. However, the situation in PBS is very different. The single-state model fails to reproduce the data (cf. SI section S-2), so an additional short state with $a_\mathrm{S} = (5.58 \pm 0.04)\,\text{\AA}$ and force constant $k_\mathrm{S} = (6 \pm 3)\,\mathrm{N/m}$ is required (blue solid line). The best-fit conformational free energy of the transition between states is $\beta\mu = -(5.0 \pm 0.4)$, indicating that the short conformation is significantly more stable.

Therefore, the short state is predominantly populated at low forces and can be identified with the helical structure proposed in previous experimental,\cite{Oesterhelt1999} molecular dynamics,\cite{Heymann1999,Liese2017} and $\textit{ab initio}$ quantum mechanical studies.\cite{Kreuzer2001c} This helical structure is originated from the formation of hydrogen-bonded bridges mediated by water molecules between next-nearest-neighboring oxygen atoms.\cite{Kreuzer2001c,Liese2017} This observation suggests that spatial correlations are lost over approximately two PEG subunit lengths (5.6\,\text{\AA}), which coincides with the Kuhn length of the short state ($a_\mathrm{S} = 5.58$\,\text{\AA}). The obtained interaction energy is very small, $\beta\varepsilon_\mathrm{SS} = (0.2 \pm 0.4)$. Therefore, the two-state model naturally reveals the absence of cooperative effects, in contrast to the model proposed in Oesterhelt \textit{et al.},\cite{Oesterhelt1999} where non-cooperativity is imposed $\textit{a priori}$.

\subsection{Negative cooperativity ($\varepsilon_\mathrm{SS}>0$): hyaluronic acid (HA)}

\begin{figure}[t]
  \centering
  \includegraphics[width=\linewidth]{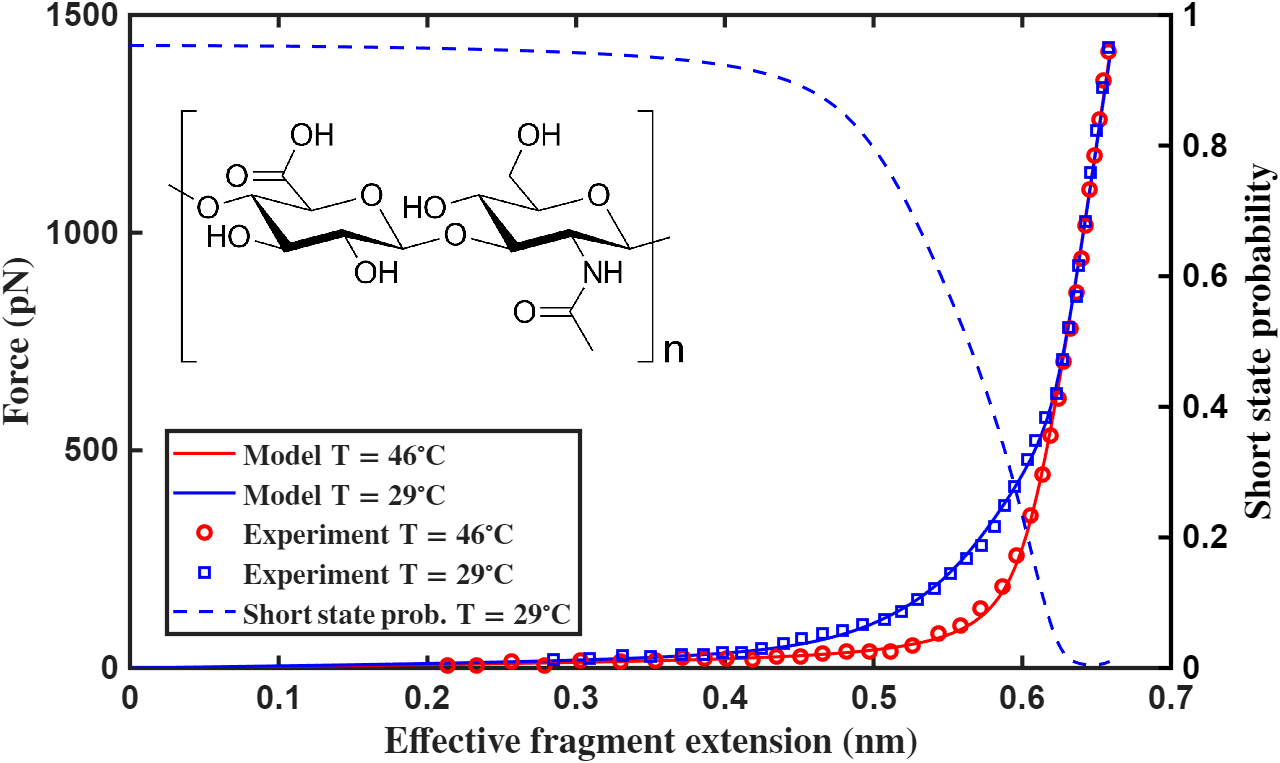}
  \caption{Experimental force–extension curves of hyaluronic acid at two different temperatures, $T = 29$\textdegree C (blue squares) and $T = 46$\textdegree C (red circles), in 0.1 M NaCl aqueous environment\cite{Giannotti2007b} shown  together with the best fits from the two-state EFJC model (solid lines). At $T = 46$\textdegree C, hydrogen bonds are broken, and the single-state EFJC model fits the data very well. In contrast, at $T = 29$\textdegree C, an additional short state must be included in the model to reproduce the data. The dashed blue line indicates the short state probability, $\theta$. The fitted parameters are: $\beta\mu = -(7.3 \pm 0.6)$, $\beta\varepsilon_\mathrm{SS} = (2.2 \pm 0.6)$, $a_\mathrm{L} = (6.0 \pm 0.2)$\,\AA, $k_\mathrm{L} = (25 \pm 2)\,\mathrm{N/m}$, $a_\mathrm{S} = (5.1 \pm 0.1)$\,\AA and $k_\mathrm{S} = (8\pm 1)\,\mathrm{N/m}$.}
  \label{fig: HA}
\end{figure}

The basic repeating unit of hyaluronic acid (HA) consists of $\beta$-(1→4)-D-glucuronic acid linked to $\beta$-(1→3)-N-acetyl-D-glucosamine (see Fig.\,\ref{fig: HA}). HA is present in the extracellular matrix\cite{Laurent1992,Toole2004} and it has important applications in medicine and cosmetics due to its remarkable mechanical properties.\cite{Necas2008} However, single-molecule stretching AFM experiments on HA remain surprisingly scarce in the literature. Yu \textit{et al.} \cite{Yu2024} compared the stretching properties of HA with those of cellulose and curdlan, relating them to the hydration properties of these molecules.\cite{Yu2024,Giannotti2007a} Here, we analyze the experimental force–extension curves reported by Giannotti \textit{et al.} \cite{Giannotti2007b} within the framework of the two-state model, since they show a temperature-dependent conformational behavior, where single HA molecules in aqueous solution acted as random coils at $T = 46$\textdegree C while superstructures were detected at $T = 29$\textdegree C.

The experimental data is shown in Fig.\,\ref{fig: HA}. Red markers correspond to temperature $T=46$\textdegree C while blue markers refer to $T=29$\textdegree C. The single-state EFJC model provides an excellent fit to the data at $T=46$\textdegree C, yielding $a_\mathrm{L}=(6.0\pm0.2)\,$\AA\, and $k_\mathrm{L}=(25\pm2)\,\mathrm{N/m}$ (see SI section S-2). However, upon decreasing the temperature to $T=29$\textdegree C an additional short state is required to reproduce the data with $a_\mathrm{S}=(5.1\pm0.1)\,$\AA\, and $k_\mathrm{S}=(8\pm1)\,\mathrm{N/m}$. The obtained energy difference between the short and the long states is $\beta\mu=-(7.3\pm0.6)$, and a conformational transition is observed at approximately 200 pN. This transition exhibits negative cooperativity, with an interaction energy between adjacent short states of $\beta\varepsilon_\mathrm{SS} = (2.2 \pm 0.6)$, which is small enough to provide similar results in the mean-field approximation (cf. SI section S-4). The pronounced sensitivity of the transition to a relatively modest increase in temperature suggests that its driving force is predominantly entropic. At $T = 46$\textdegree C, the increase in configurational entropy offsets the stabilizing effect of hydrogen bonding, thereby reducing the free-energy difference between the two conformational states.

The existence of short states in HA has been previously reported in the literature and has been attributed to the presence of intramolecular hydrogen bonds,\cite{Heatley1988,Giannotti2007b} which are disrupted under applied force. X-ray diffraction studies have provided evidence for helical structures involving two, three, or four monomer units.\cite{Sheehan1975,Winter1975,Guss1975} Ordered structures stabilized by hydrogen bonding, as well as helix–coil transitions in solution, have also been described.\cite{Milas2004,Hirano1973,Ghosh1993} Similar behavior has been observed in other polysaccharides. Li \textit{et al.}\cite{Li1999} and Marszalek \textit{et al.} \cite{Marszalek1998} reported a force-induced chair-boat transition in $\alpha$-(1,4) polysaccharides such as carboxymethyl amylose (CM-amylose) and heparin, whereas the transition is absent in $\beta$-(1,4) polysaccharides, such as CM-cellulose. By analyzing separately the low- and high-force regimes of the force-extension curve and fitting each region independently to a single-EFJC model, Li \textit{et al.} \cite{Li1999} determined Kuhn lengths ($a_\mathrm{L}=6.2\,$\AA\,,\,$a_\mathrm{S}=5.4\,$\AA), elastic constants ($k_\mathrm{L}=28$\,N/m,\,$k_\mathrm{S}=11$\,N/m) and a transition free energy of $\beta\mu=-(7.3\pm0.6)$. Cooperativity was not considered in that work. These values are strikingly similar to those obtained here for HA. However one should be cautious comparing the results for CM-amylose to those of HA, a considerably more complex molecule. In HA the glycosidic linkages are of the $\beta$-(1,4) and $\beta$-(1,3) types, unlike those in CM-Amylose, which are of $\alpha$-(1,4). Moreover, the presence of amide groups broadens the possible configurations able to form hydrogen bonds. Further studies, possibly involving enhanced simulations with explicit solvent, would be necessary and desirable to elucidate the detailed mechanism of the transition in HA.

\subsection{Positive cooperativity ($\varepsilon_\mathrm{SS}<0$): B-DNA to S-DNA conformational transition}

\begin{figure}[b]
  \centering
  \includegraphics[width=\linewidth]{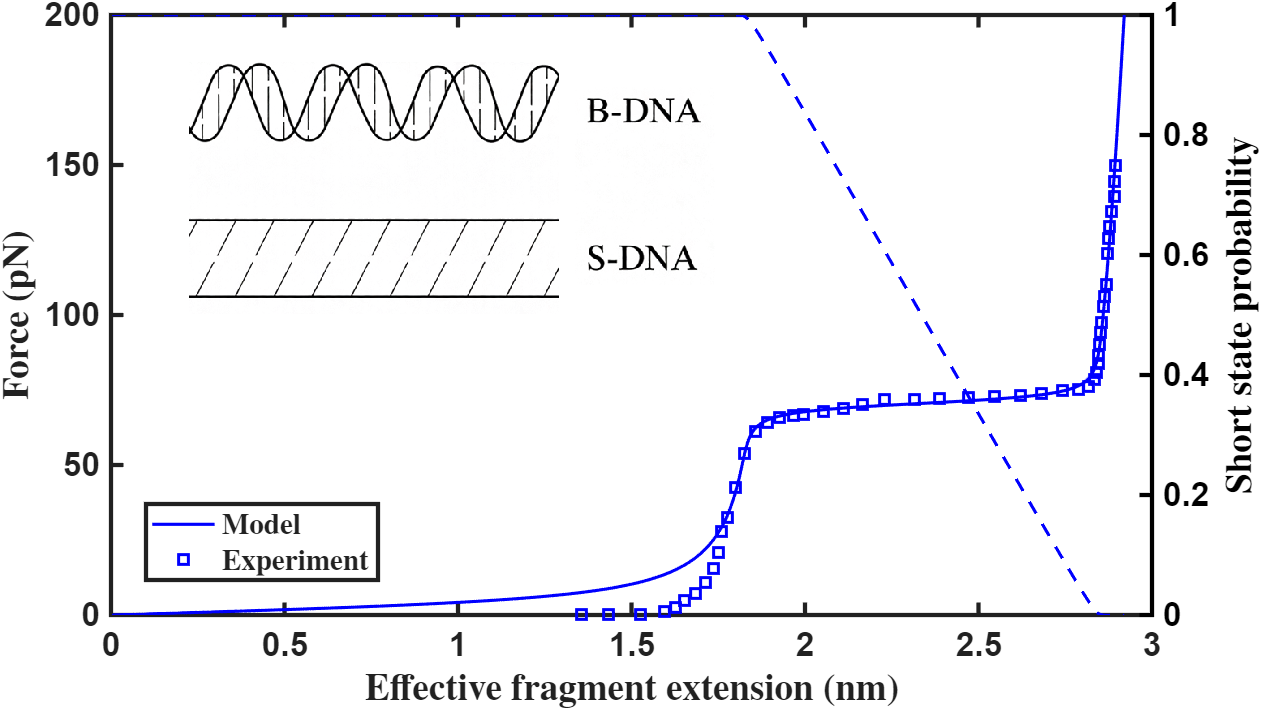}
  \caption{Experimental force-extension curve of DNA molecule in PBS solution (100 mM; 80 mM Na$^+$ and 0.01\% Tween) (blue markers)\cite{Cluzel1996} with the associated best-fit two-state model (solid line).  The dashed blue line represents the short state probability, $\theta$, which clearly resembles the one expected from the lever rule in phase equilibrium. The fitted parameters are $\beta\mu=-14.2$, $\beta\varepsilon_\mathrm{SS}=-2.0$, $a_\mathrm{L}=28.6\,$\AA $\;$, $k_\mathrm{L}=4\,\mathrm{N/m}$ and $a_\mathrm{S}=19\,$\AA. In this case, likely due to the steep behavior of the force–extension curve across the transition, the Levenberg–Marquardt algorithm did not converge reliably. Therefore, the parameters were determined by fixing some of them based on reasonable physical criteria: $k_\mathrm{L}$ is determined by measuring the slope at forces larger than 90 pN whereas the B-DNA form is assumed to be effectively rigid ($k_\mathrm{S} \to \infty$). Equation \ref{eq: transition general condition} is then used to constrain the parameter space compatible with the sharp transition.}
  \label{fig: DNA}
\end{figure} 
Because of its biological importance and well-known structure, the DNA molecule has been extensively used as a model system in single-molecule stretching experiments. When subjected to forces of about 65 pN, \cite{Smith1996,Mameren2009} DNA undergoes an abrupt conformational transition from the canonical (or relaxed) B-DNA form to a new structure known as S-DNA, in which the base pairs are more widely separated. As a result, the Kuhn length increases by a factor of approximately 1.7.\cite{Cluzel1996,Smith1996,Mameren2009}

Figure\,\ref{fig: DNA} shows the experimental force–extension curve obtained by Cluzel \textit{et al.} \cite{Cluzel1996} (markers), together with the best fit of the two-state model (solid line). The dashed line indicates the probability of the short B-DNA state, whose roughly linear behavior can be interpreted using the lever rule of phase equilibrium.\cite{Smith1996} Considering that the short B-DNA fragment corresponds to half a canonical helix of approximately five base pairs,\cite{Wang1979,Dickerson1981,Ahsan1998} the fitted values of $a_\mathrm{S}$ and $a_\mathrm{L}$ are close to the expected 3.4\,\AA\ and 5.8\,\AA\ per base pair reported in previous works.\cite{Wenner2002,Smith1996} In Smith \textit{et al.},\cite{Smith1996} the transition free energy was estimated as 9.3 kJ/mol per base pair, corresponding to 3.72 $k_\mathrm{B}T$ at $T = 25^\circ$C. Using the two-state model, the transition free energy can be approximated as the elimination of one $\mu$ and two $\varepsilon_\mathrm{SS}$ terms per fragment. Dividing by five base pairs per fragment yields a free energy of roughly 3.66\,$k_\mathrm{B}T$ per base pair, in good agreement with the experimental estimate.

We observe that the model successfully captures the conformational transition, which, despite DNA being effectively one-dimensional, resembles a first-order phase transition.\cite{Smith1996}. In this case, although the model employed here is exact, we also find that the mean-field approximation still provides an excellent description, despite being close to the limits of its applicability (see Fig. \ref{fig:comparison MF} and S.I.). At high forces, the S-state behaves similarly to single-stranded DNA and is therefore well described by the EFJC model,\cite{Smith1992} with an elasticity constant $k_\mathrm{L}$ comparable to the values reported by Wenner \textit{et al.} \cite{Wenner2002} However, below 50 pN, the fit deteriorates significantly, reflecting the limitations of the EFJC approximation. In this regime, DNA seems to be more accurately described by an elastic worm-like chain (EWLC) model.\cite{Cluzel1996,Ahsan1998} We suggest that the two-state framework could be refined to include multiple stretching “modes”—that is, the simultaneous presence of EFJC and EWLC fragments— to better capture such mixed mechanical responses, as proposed in Jarillo.\cite{Jarillo2017}

\section{\label{sec: transitions} Conditions and mechanisms of Sharp Conformational Transitions}

Sharp conformational transitions are important because they confer switch-like behavior to the system: small changes in the control variables can induce drastic changes in the macromolecular structure. 
This situation is common in biological systems, with the cooperative behavior of hemoglobin being a paradigmatic example.\cite{Wyman1990,Acerenza1997,Garces2008}

We analyze the conditions under which the two-state EFJC model exhibits sharp conformational transitions at a critical force $f_{c}$. Below $f_{c}$, chain fragments prevail in a certain state, whereas above $f_{c}$ the other state becomes predominant. If such a critical force exists, it corresponds approximately  to the solution of the second-order equation

\begin{gather}
\frac{\beta f_{\mathrm{c}}^2}{2}\left(\frac{1}{k_1}-\frac{1}{k_0}\right)
+ \notag \\
+\beta\left(a_1-a_0\right)f_\mathrm{c}
=\beta\left(\varepsilon+\mu\right)
+\ln\left(\frac{a_1}{a_0}\right)
\label{eq: transition general condition}
\end{gather}
where $f_{\mathrm{c}}$ must also satisfy $f_{\mathrm{c}} < a_\alpha k_\alpha$ (with $\alpha=0,1$) to avoid nonphysical transitions at forces where the harmonic potential is no longer valid. 
A detailed derivation can be found in section S-5 in the SI.

\begin{figure}[b]
    \centering
    \includegraphics[width=1\linewidth]{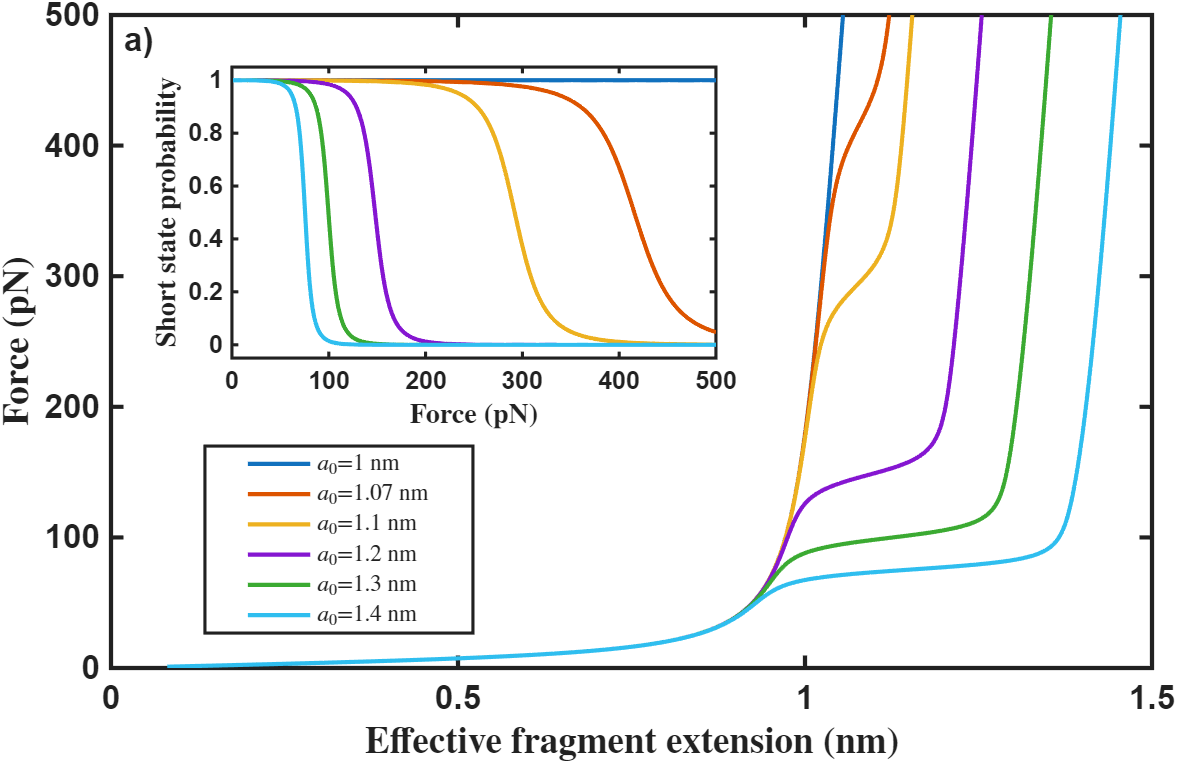}
    \includegraphics[width=1\linewidth]{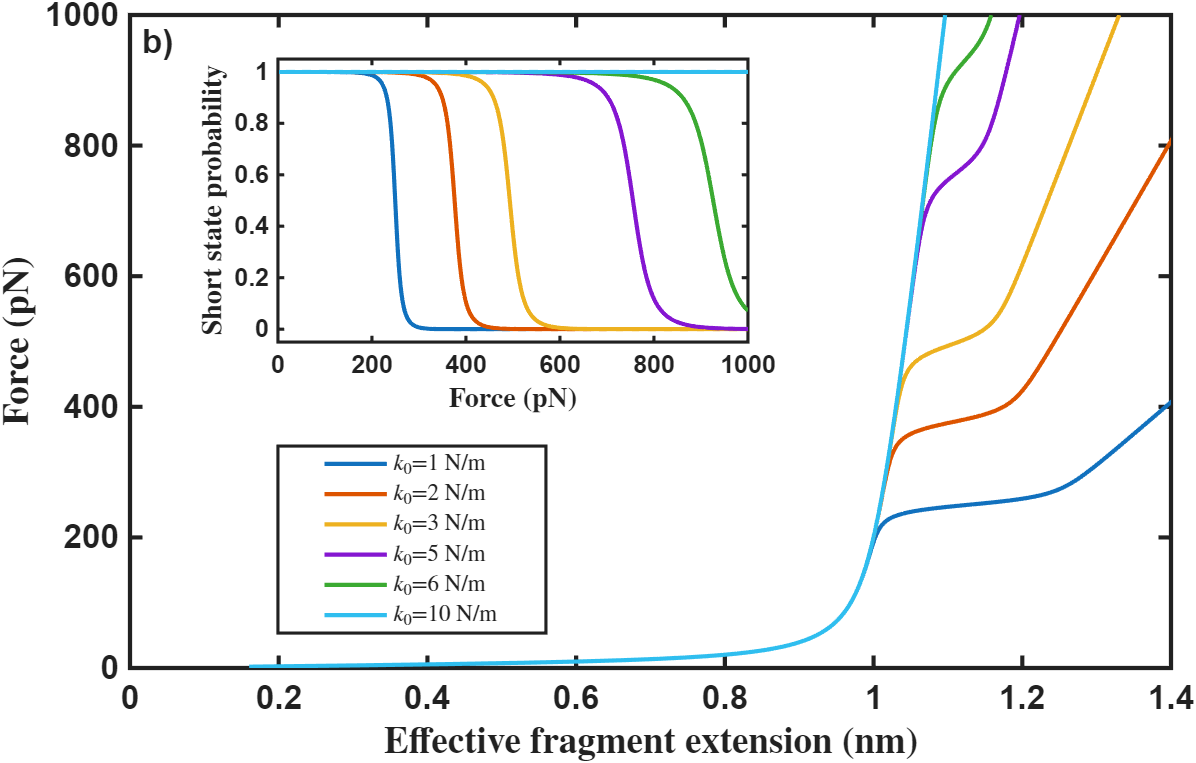}
    \caption{Force–extension curves exhibiting conformational transitions induced by: (a) differences in fragment lengths, with $k_0 = k_1 = 8$ N/m, $a_1 = 1$ nm, and $a_0$ = 1, 1.07, 1.1, 1.2, 1.3, 1.4 nm; (b) differences in elastic constants, with $a_0 = a_1 = 1$ nm, $k_1 = 10$ N/m, and $k_0$ = 1, 2, 3, 5, 6, 10 N/m. The remaining parameters are $T = 298$ K, $\beta\mu = -5$, and $\beta\varepsilon_\mathrm{SS} = -2$. The inset plots show the state-one probability $\theta$ as a function of force, $f$.}
    \label{fig:Transitions}
\end{figure}

An interesting consequence of Eq.\,\ref{eq: transition general condition} is that sharp conformational transitions can arise from two distinct mechanisms or, more generally, from a combination of both. Setting $k_0=k_1$ in Eq.\,\ref{eq: transition general condition} yields the condition
\begin{equation}
f_\mathrm{c}\approx\frac{\mu+\varepsilon+\beta^{-1}\ln\left(\frac{a_1}{a_0}\right)}{a_1-a_0} 
\label{eq:transition force FJC as}
\end{equation}
which does not depend on the elastic constants. Without loss of generality, let us assume that $a_1<a_0$. Since $f_\mathrm{c}$ is positive, the numerator must be negative: $\mu+\varepsilon+\beta^{-1}\ln\left(\frac{a_1}{a_0}\right)<0$. This condition implies that, for forces below $f_{\mathrm{c}}$ the short state is thermodynamically favored. The long state is separated by a free-energy barrier determined by $\mu$, $\varepsilon$ and the entropic contribution $\beta^{-1}\ln\left(\frac{a_1}{a_0}\right)$. The transition occurs when the mechanical work $f_\mathrm{c}\left(a_1-a_0\right)$ compensates for this barrier (see Fig.\,\ref{fig:Transitions}a). This first mechanism therefore requires the two conformational states to have different Kuhn lengths.

A second possible mechanism emerges when considering the limiting case in which $a_0=a_1$ but $k_0\neq k_1$. In this situation, the solution of Eq.\,\ref{eq: transition general condition} reads
\begin{equation}
f_\mathrm{c}\approx\sqrt{\frac{2\left(\mu+\varepsilon\right)}{\frac{1}{k_1}-\frac{1}{k_0}} }
\label{eq:transition force FJC ks}
\end{equation}
which is independent of $a_0$ and $a_1$. In this case, the transition is driven by the difference in elastic constants between the two conformations. For instance, ligand binding may alter the chain stiffness without significantly affecting the fragment length.
For $f_\mathrm{c}$ to be a real number, numerator and denominator under the square root must have identical signs. 
Supposing that $k_1>k_0$, the preceding statement is hold  only if $\mu+\varepsilon<0$, and the state-1 stiffer conformation is the most populated at zero force. Therefore the transition takes place when the system overcomes the $\mu+\varepsilon$ barrier and is more easily stretched in the 0-state conformation due to the lower value of $k_0$ compared to $k_1$ (see Fig.\,\ref{fig:Transitions}b). In the general case, both mechanisms may act simultaneously or even cooperate to induce the conformational transition.

\section{\label{sec: Generalization to more than two conformational states}Generalization to more than two conformational states}

We now generalize the previous discussion to an arbitrary number,  $M$, of conformational states. 
Such a situation arises, for example, when two or more ligands compete for the same binding sites along the chain. In this case, the free energy given in Eq.\,\ref{eq: global elastic free energy} must be rewritten as follows.\cite{Garces2006} 
Suppose that the $i$-th fragment is in a given state $\nu$ ($\nu=1,\dots,M$). 
The state of such fragment can be represented by a "state vector" $\mathbf{s}_i$ with $M$ components, all zero except for the $\nu$-th component, which equals one, 
\begin{equation}
\mathbf{s}_i = (0,\;\ldots,\;0,\;\overset{(\nu)}{1},\;0,\;\ldots,\;0).
\label{eq: occupation vector}
\end{equation}

For instance, when $M=3$ the vector $\mathbf{s}_4=\left(0,1,0\right)$ indicates that the fourth fragment $(i=4)$ is in state $\nu=2$. With this notation, the free energy can be expressed as
\begin{equation}
    F\left(\{\mathbf{s}_i, \mathbf{b}_i\}\right)=-\sum_{i=1}^N \mathbf{f}\cdot\mathbf{b}_i+\sum_{i=1}^N \mathbf{e}_i\cdot\mathbf{s}_i^\mathrm{T} + \sum_{i=1}^{N-1} \mathbf{s}_i\mathbf{E}\,\mathbf{s}_{i+1}^\mathrm{T}
    \label{eq: global elastic free energy M conformations}
\end{equation}
where $\mathbf{e}_i$ is a vector containing the transition free energy and elastic contributions for each state
\begin{equation}
    \left(\mathbf{e}_i\right)_\nu=\mu_\nu+\frac{k_\nu}{2}\left(b_{i}-a_\nu\right)^2
    \label{eq: vector of energies}
\end{equation}
and $\mathbf{E}$ is a matrix whose elements encode the interaction energies between neighboring fragments in states $\nu$ and $\gamma$, i.e. $\left(\mathbf{E}\right)_{\nu\gamma}=\varepsilon_{\nu\gamma}$. Note that this vector notation also describes the two-state model, but it was not introduced earlier for simplicity. 
Proceeding in the same way as in section \ref{sec: EXACT SOLUTION FOR THE TWO-STATE FREELY JOINTED CHAIN} and using a similar trick as in Eq.\,\ref{eq: conformational g's def} the integrals over the fragment vectors can be evaluated. As a result an effective free energy is obtained
\begin{gather}
    F_{\mathrm{eff}}\left(\{\mathbf{s}_i\}\right)=Ng_1+
    \sum_{i=1}^N \mathbf{\tilde{\mu}}\cdot\mathbf{s}_i^\mathrm{T} + \sum_{i=1}^{N-1} \mathbf{s}_i\mathbf{E}\,\mathbf{s}_{i+1}^\mathrm{T}
    \label{eq:effective free energy for M states}
\end{gather}
where the components of the vector $\tilde{\mu}$ read
\begin{gather}
\tilde{\mu}_\nu\left(f\right)=g_\nu\left(f\right)-g_1\left(f\right)+\mu_\nu,
    \label{eq:effective energy vector for M states}
\end{gather}
taking $\nu=1$ as the reference state. 
The expression from Eq.\,\ref{eq:effective free energy for M states} corresponds to the free energy of a Potts model, \cite{Baxter1982} a generalization of the Ising model to more than two states, and its partition function (Eq.\,\ref{eq: global effective part func}) can be calculated using transfer matrices. 
The full derivation, provided in the SI section S-6, parallels the two-state case, and Eq.\,\ref{eq:pt single ligand effective} remains valid when applied with the corresponding transfer matrix. 
The $\nu$-state probability, $\theta_\nu$, as well as the effective fragment extension, can now be obtained from the partition function as
\begin{gather}
\theta_\nu=-\frac{1}{\beta N}\frac{\partial\ln{\Xi}}{\partial\mu_\nu}\;\;\; \text{and}\;\;\;
x=\sum_{\nu=1}^{M}\theta_\nu\;x_\nu
\label{eq: occupation number and relative elongation M states}
\end{gather}
where $x_\nu$ denotes the effective extension of the fragment in state $\nu$ along the direction of the applied force.

\begin{figure}[b]
  \centering
  \includegraphics[width=\linewidth]{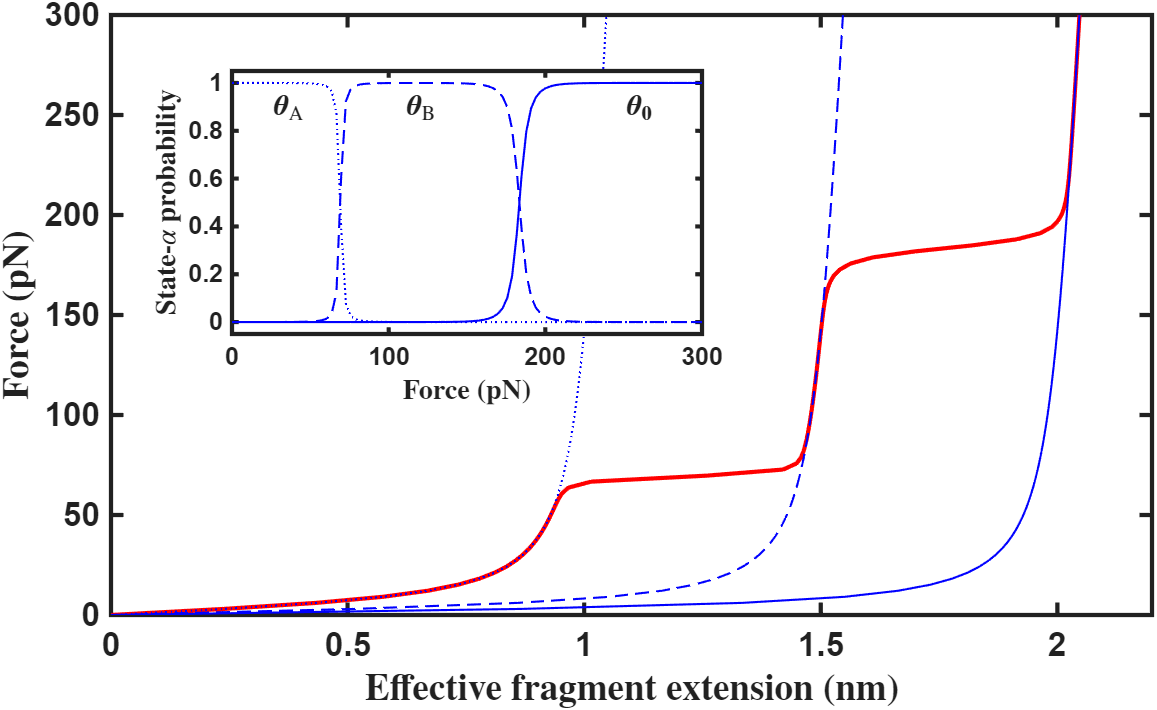}
  \caption{Force–extension ($f-x$) curve for a ligand–receptor system in which two ligands (A and B) compete for binding to the same receptors. Three states are therefore possible: unbound, bound to ligand A, and bound to ligand B (denoted by $\emptyset$, A, and B, respectively). The parameter values are $T = 298\,\mathrm{K}$, $a_\mathrm{A} = 1\,\mathrm{nm}$, $a_\mathrm{B} = 1.5\,\mathrm{nm}$, $a_\emptyset = 2\,\mathrm{nm}$ and $k_\mathrm{A} = k_\mathrm{B} = k_\emptyset = 5\,\mathrm{N/m}$, with $\beta\mu_{\mathrm{A}} = -28$, $\beta\mu_{\mathrm{B}} = -20$, $\beta\varepsilon_{\mathrm{AA}} = \beta\varepsilon_{\mathrm{BB}} = -2$, and $\beta\varepsilon_{\mathrm{AB}} = 0$. The inset shows the probabilities $\theta$ of states $\emptyset$, A, and B as functions of the applied force, $f$.}
  \label{fig: Competition}
\end{figure}

For instance, consider a ligand–receptor system in which two ligands, A and B, compete for binding to the same receptor. The receptor can exist in three possible states: unbound (denoted by $\emptyset$, $\nu=1$), bound to ligand A ($\nu=2$), or bound to ligand B ($\nu=3$). Therefore, the required transfer matrix in this case reads
\begin{equation}
\tilde{\mathbf{T}}=\left(\begin{array}{ccc}
1 & \tilde{z}_\mathrm{A} & \tilde{z}_\mathrm{B}\\
1 & \tilde{z}_\mathrm{A}u_\mathrm{AA} & \tilde{z}_\mathrm{B}u_\mathrm{AB}\\
1 & \tilde{z}_\mathrm{A}u_\mathrm{AB} & \tilde{z}_\mathrm{B}u_\mathrm{BB}\\
\end{array}\right)
\label{eq: transfer matrix competitive}
\end{equation}
Then, the partition function is obtained by substituting the maximum eigenvalue of this transfer matrix into Eq.\,\ref{eq: part func infinity} and the force–extension curves are calculated using Eq.\,\ref{eq: occupation number and relative elongation M states} and displayed in Fig.\,\ref{fig: Competition}. At zero force, the receptors are predominantly occupied by ligand A. Upon increasing the force to approximately 70 pN, a transition to ligand B binding takes place, giving rise to a first plateau in the extension curve. As the force is further increased to about 170 pN, ligand B dissociates and the receptors become unbound. In this regime, mechanical force effectively acts as a switch that selectively promotes or suppresses binding to a given ligand. Such force-controlled selectivity might open interesting possibilities for nanotechnological applications.

\section{\label{sec:conclusions}Conclusions}

In this work, we investigate the stretching behavior of linear macromolecules that undergo cooperative conformational transitions, whether induced by the binding of external ligands or arising from intrinsic intramolecular equilibrium.

We present an alternative derivation\cite{Benedito2018} of the two-state Elastic Freely Jointed Chain (EFJC) model originally introduced by Radiom and Borkovec.\cite{Radiom2017a} This minimal model incorporates the smallest set of parameters required to describe a transition between two conformational states: two Kuhn lengths, two elastic constants, a free energy difference between states, and an interaction energy accounting for cooperativity. In the thermodynamic limit of long chains, useful and simple expressions for both the fragment probability and the effective fragment extension are provided, leading to a straightforward analytical treatment of polymer stretching experiments.

The model offers an excellent fit to the force–extension curves of poly(ethylene-glycol) (PEG) and hyaluronic acid (HA) from reported AFM-SMFS experiments.\cite{Oesterhelt1999,Giannotti2007b} No cooperativity is observed during the stretching of PEG. In contrast, negative cooperativity is observed in HA during the transition between both states. These results are consistent with previous studies highlighting the role of hydrogen bonding in the conformations of both PEG and HA. Finally, the model is applied to describe the force-induced B-DNA to S-DNA conformational transition,\cite{Cluzel1996} where positive cooperativity is observed.

We also analyze the mathematical characteristics of sharp conformational transitions. Such transitions can be driven by two distinct mechanisms: a difference in Kuhn lengths or a difference in elastic constants. These mechanisms may act independently, cooperate, or even compete.

Finally, we discuss the extension of the model to systems with more than two states. For example, a chain fragment may be unoccupied or bound by one of two possible ligands. It is shown that the applied force can modulate the system’s preference for one ligand over the other, effectively acting as a switch that controls binding selectivity.

\section{\label{sec:supplementary material}Supplementary Material}
This article is accompanied by Supplementary Material:
\begin{itemize}
    \item A Supplementary Information document containing additional derivations that were omitted from the main text for clarity and space considerations.
    \item A zip folder \textit{SI\_Fitting\_HA}  containing the MATLAB code used for the fitting protocol and the reported HA data.\cite{Giannotti2007b}
\end{itemize}

\section{\label{sec:authors contribution}Authors' Contribution}
All authors have contributed to this work and declare that they have no conflict of interest.

\section{\label{sec:aknowledgements} Acknowledgements}
P.M.B., S.M. and F.M. acknowledge the financial support from Generalitat de Catalunya (Grant 2021SGR00350). S.M. and F.M. acknowledge Spanish Structures of Excellence María de Maeztu program through Grant CEX2021-001202-M. M.I.G thanks the Agencia Estatal de Investigación for the financial support through the research project PID2022-140459OB-I00. P.M.B  and J.L.G. also thank the same institution through the research projects PID2022-140312NB-C21 and PID2024-156219NB-C21. P.M.B. has received funding from the postdoctoral fellowships programme Beatriu de Pinós (2024BP00134), funded by the Secretary of Universities and Research (Government of Catalonia) and by the Horizon 2020 programme of research and innovation of the European Union under the Marie Sklodowska-Curie grant agreement No 801370.
J.O. acknowledges the financial support of the FPU grant from the Spanish Ministry of Innovation, Science and Universities (FPU21/05318).

\section{\label{sec:data av.stat.}Data Availability Statement}
The data that support the findings of this study are available from the author (J.O. at j.orradre@ub.edu) upon reasonable request.

\section{\label{sec:References} References}

\bibliography{bibliography}

\end{document}